\documentclass[twoside,a4paper,fleqn]{article}%
\usepackage[headings]{espcrc2}
\usepackage{graphicx}
\usepackage[figuresright]{rotating}
\usepackage{amsmath}%
\usepackage{amsfonts}%
\usepackage{amssymb}
\readRCS
$Id: espcrc2.tex,v 1.2 2004/02/24 11:22:11 spepping Exp $
\runtitle{Theory review on rare K decays: SM and beyond}
\runauthor{Christopher Smith}

\begin{document}

\title{Theory review on rare K decays: Standard Model and beyond}
\author{Christopher Smith\\Institut f\"{u}r Theoretische Physik, Universit\"{a}t Bern, CH-3012 Bern,
Switzerland \thanks{Work supported by the Schweizerischer Nationalfonds.}}

\begin{abstract}
The theoretical status of the rare $K\rightarrow\pi\nu\bar{\nu}$,
$K_{L}\rightarrow\pi^{0}\ell^{+}\ell^{-}$ and $K_{L}\rightarrow\mu^{+}\mu^{-}$
decays in the Standard Model is reviewed. Their sensitivity to New Physics and their
discriminating power is also illustrated. \vspace{0.9pc}
\end{abstract}
\maketitle


\section{Rare K decays in the Standard Model}

One of the reasons why rare K decays are interesting is the good
theoretical control reached over their predictions in the SM. In this section,
the ingredients needed are shortly reviewed.

\subsection{Electroweak structure}

The electroweak processes driving the rare semi-leptonic K decays are the $W$
box, $Z$ and $\gamma$ penguins\cite{BuchallaBL96}, and lead to the amplitude
\[
\mathcal{A}\left(  K_{L}\rightarrow\pi^{0}X\right)  =\sum_{q=u,c,t}\left(
\operatorname{Im}\lambda_{q}+\varepsilon\operatorname{Re}\lambda_{q}\right)
y_{q}^{X}%
\]
with $X=\nu\bar{\nu},\ell^{+}\ell^{-}$, $\lambda_{q}=V_{qs}^{\ast}V_{qd}$.
Without the dependence of the loop functions $y_{q}^{X}$ on the quark
masses, CKM unitarity would imply vanishing FCNC (GIM mechanism). Now, looking
at these dependences, combined with the scaling of the CKM elements, one can
readily get a handle on the importance of each quark contribution.

For $X=\nu\bar{\nu}$, only the $Z$ penguin and $W$ box enter, $y_{q}^{\nu
\bar{\nu}}\sim m_{q}^{2}$, and light quark contributions are suppressed.
Since, in addition, $\varepsilon\sim10^{-3}$ and $\operatorname{Re}\lambda
_{t}\sim\operatorname{Im}\lambda_{t}$, indirect CP-violation is small. For
$K^{+}$, the $\operatorname{Re}\lambda_{c,t}$ and $\operatorname{Im}\lambda
_{t}$ parts contribute.

For $X=\ell^{+}\ell^{-}$, the photon penguin enters with its scaling
$y_{q}^{\ell\ell}\sim\log(m_{q})$ for $m_{q}\rightarrow0$. Direct CP-violation
is still short-distance dominated thanks to $\operatorname{Im}\lambda_{u}=0$,
but not indirect CP-violation, completely dominated by the long-distance
$u$-quark photon penguin, $K_{1}\rightarrow\pi^{0}\gamma^{\ast}\rightarrow
\pi^{0}\ell^{+}\ell^{-}$.

For $K_{L}\rightarrow\ell^{+}\ell^{-}$, the structure is similar to
$K\rightarrow\pi\nu\bar{\nu}$, up to the change $\operatorname{Im}\lambda
_{q}\leftrightarrow\operatorname{Re}\lambda_{q}$ (no single-photon penguin).

Finally, for charged leptons, there is also the double-photon penguin, which
gives a CP-conserving contribution ($\sim\operatorname{Re}\lambda_{q}$) to
$K_{L}\rightarrow\pi^{0}\ell^{+}\ell^{-}$ and $K_{L}\rightarrow\ell^{+}%
\ell^{-}$, and is completely dominated by long-distance ($u$-quark).

\subsection{QCD corrections}

Having identified the relevant electroweak structures, both perturbative and
non-perturbative QCD effects have now to be included. This is done in three
main steps:

\textit{Step 1}: Integration of heavy degrees of freedom (top, W, Z),
including perturbative QCD effects above $M_{W}$. This generates local FCNC
operators, and Fermi-type four fermion local operators.

\textit{Step 2}: Resummation of QCD corrections (running down). At the $c$
threshold (similar for $b,\tau$), four-fermion operators are combined to form
closed $c$ loops, which are then replaced by a tower of effective interactions
in increasing powers of (external momentum)/(charm mass). The lowest order
consists again of the dim.6 FCNC operators, while dim.8 operators are
corrections scaling naively like $m_{K}^{2}/m_{c}^{2}\sim15\%$.

\textit{Step 3}: To get the amplitude, it remains to compute the matrix
elements of all the operators obtained. Those of dim.6 semi-leptonic
operators can be extracted from $K_{\ell2},K_{\ell3}$ decays (with Chiral
Perturbation Theory (ChPT) corrections). Contributions from four-quark
operators are represented directly in terms of meson fields in ChPT, such that
non-local $u$-quark loops are represented as meson loops. The price to pay are
some unknown low-energy constants, to be extracted from experiment. For dim.8
operators, an approximate matching is done with the ChPT representation of the
$u$-quark contributions.

\subsection{The $K\rightarrow\pi\nu\bar{\nu}$ decays in the SM}

A high level of precision is attained for these modes. Dim.6 FCNC
operators from the $t$-quark are known at NLO, while $c$-quark ones have
recently been obtained at NNLO\cite{BurasNNLO}. The matrix elements for these
operators are known from $K_{\ell3}$, including the leading isospin corrections
\cite{MP}. For $K^+\rightarrow\pi^+\nu\bar{\nu}$, residual $c$-quark effects
from dim.8 operators, along with long-distance $u$-quark effects have also been
computed\cite{IsidoriMS}. For $K_{L}\rightarrow\pi^{0}\nu\bar{\nu}$,
ICPV is of about 1\%\cite{BB96} while the CP-conserving contribution is less
than 0.01\%\cite{BI98}.

The SM predictions are $\mathcal{B}\left(  K_{L}\rightarrow\pi^{0}\nu\bar{\nu
}\right)  =(2.81\pm0.56)\cdot10^{-11}$ and $\mathcal{B}\left(  K^{+}%
\rightarrow\pi^{+}\nu\bar{\nu}\right)  =(8.0\pm1.1)\cdot10^{-11}$. The error
on $K_{L}\rightarrow\pi^{0}\nu\bar{\nu}$ is dominated by $\operatorname{Im}%
\lambda_{t}$, while for $K^{+}\rightarrow\pi^{+}\nu\bar{\nu}$, which receives
a significant $c$-quark contribution, it breaks down to\cite{BurasNNLO} scales
(13\%), $m_{c}$(22\%), CKM, $\alpha_{S}$, $m_{t}$ (37\%) and matrix-elements
from $K_{\ell3}$ and light-quark contributions (28\%). Further improvements
are thus possible through a better knowledge of $m_{c}$, of the isospin
breaking in the $K\rightarrow\pi$ form-factors, or by a lattice study of
higher-dimensional operators.

\subsection{The $K_{L}\rightarrow\pi^{0}\ell^{+}\ell^{-}$ decays in the SM}

Here the situation is more complicated. The $t$ and $c$ quark contributions,
known to NLO, generate both the dimension-six vector $(\bar{s}d)_{V}(\ell\ell)_{V}$
and axial-vector $(\bar{s}d)_{V}(\ell\ell)_{A}$ operators. The former produces
the $\ell^{+}\ell^{-}$ pair in a $1^{--}$ state, the latter in both $1^{++}$
and $0^{-+}$ states.

Indirect CP-violation is related to $K_{S}\rightarrow\pi^{0}\ell^{+}\ell^{-}$,
dominated by the ChPT counterterm $a_{S}$\cite{DEIP}. NA48 measurements give
$\left|  a_{S}\right|  =1.2\pm0.2$\cite{KSpill}. Producing $\ell^{+}\ell^{-}$
in a $1^{--}$ state, it interferes with the contribution from the $(\bar
{s}d)_{V}(\ell\ell)_{V}$ operator, arguably constructively\cite{BDI03,FGD04}.

The CP-conserving contribution from $Q_{1,...,6}$ proceeds through
two-photons, i.e. produces the lepton pair in either a helicity-suppressed
0$^{++}$ or phase-space suppressed $2^{++}$ state. The LO corresponds to the
finite two-loop process $K_{L}\rightarrow\pi^{0}P^{+}P^{-}\rightarrow\pi
^{0}\gamma\gamma\rightarrow\pi^{0}\ell^{+}\ell^{-}$, $P=\pi,K$, exactly
predicted by ChPT, and produces only 0$^{++}$ states. Higher order corrections
are estimated using experimental data on $K_{L}\rightarrow\pi^{0}\gamma\gamma$
for both the $0^{++}$ and $2^{++}$ contributions\cite{BDI03,IsidoriSmithUnter}.

Altogether, the predicted rates are $\mathcal{B}_{\mathrm{SM}}^{e^{+}e^{-}%
}=3.54_{-0.85}^{+0.98}\;\left( 1.56_{-0.49}^{+0.62}\right) \cdot10^{-11}$
and $\mathcal{B}_{\mathrm{SM}}^{\mu^{+}\mu^{-}}=1.41_{-0.26}^{+0.28}\;\left(
0.95_{-0.21}^{+0.22}\right)  \cdot10^{-11}$ for constructive (destructive)
interference. The errors are detailed in \cite{BDI03,IsidoriSmithUnter,MesciaST}. 
Overall, the error on $a_{S}$ is currently the most limitative and better
measurements of $K_{S}\rightarrow\pi^{0}\ell^{+}\ell^{-}$ would be welcomed.

Finally, the integrated forward-backward (or lepton-energy) asymmetry (see
Refs in \cite{MesciaST}) is generated by the interference between
CP-conserving and CP-violating amplitudes. While for $A_{FB}^{e}$, no reliable
prediction can be made because of the poor theoretical control on the $2^{++}$
contribution, the situation is better for $A_{FB}^{\mu}$, for which the
$0^{++}$ contribution is under control. Though the error is large, it can be
used to fix the interference sign since $A_{FB}^{\mu}\approx-25\%$ or $15\%$
depending on sign$(a_{S})$.

\subsection{The $K_{L}\rightarrow\mu^{+}\mu^{-}$ decay in the SM}

The short-distance (SD) piece from $t$ and $c$-quarks is known to NLO and
NNLO\cite{GorbahnHaisch}, resp., and is helicity-suppressed. Indirect
CP-violation is negligible. The long-distance (LD) contribution, from the
matrix elements of $Q_{1,...,6}$, proceeds again through two-photons. Still,
there are three differences with respect to $K_{L}\rightarrow\pi^{0}\ell
^{+}\ell^{-}$.

First, the contribution from the imaginary part of the photon loop, estimated
from $K_{L}\rightarrow\gamma\gamma$, is much larger than SD, and already accounts 
for the bulk of the experimental $K_{L}\rightarrow\mu^{+}\mu^{-}$ rate.
Second, while the charged meson loop in $K_{L}\rightarrow\pi^{0}P^{+}%
P^{-}\rightarrow\pi^{0}\gamma\gamma\rightarrow\pi^{0}\ell^{+}\ell^{-}$ is
acting like a cut-off, and a finite result is found, now the two photons arise
from the axial anomaly, and $K_{L}\rightarrow\pi^{0},\eta,\eta^{\prime
}\rightarrow\gamma\gamma\rightarrow\mu^{+}\mu^{-}$ is divergent. Though still
with a large theoretical error, the dispersive $\gamma\gamma$ part was
estimated using experimental information on $K_{L}\rightarrow\gamma^{\ast
}\gamma^{\ast}$ and the perturbative behavior of the $\bar{s}d\rightarrow
\bar{u}u\rightarrow\gamma\gamma$ loop\cite{IsidoriU03}. Finally, both SD and
LD produce the lepton pair in the same 0$^{-+}$ state and interfere with an
unknown sign, which depends on that of $\mathcal{A}(K_{L}\rightarrow
\gamma\gamma)$. In this respect, the progress made in \cite{GerardST} for
treating $K_{L}\rightarrow\gamma\gamma$ points towards constructive
interference between SD and LD. As shown there, confirmation of this sign
could be obtained from better measurements of $K_S\rightarrow\pi^0%
\gamma\gamma$ or $K^+\rightarrow\pi^+\gamma\gamma$.

\section{New Physics in rare K decays}

Being suppressed in the standard model, and in addition, the SM predictions
being under theoretical control, makes the rare $K$ decays ideal to get clear
signals of New Physics (NP). Even if LHC finds NP signals before Kaon
experiments, it will remain essential to probe the $\Delta S=1$ sector.
Indeed, in general, NP models involve many new flavor breaking parameters.
Experimental information will be necessary to establish their structure, and
thereby, give us some hints about a possible higher level of unification.

\subsection{New Physics in $K\rightarrow\pi\nu\bar{\nu}$ decays}

Model-independently, the present measurement of $\mathcal{B}\left(K^+\rightarrow\pi^+\nu\bar{\nu}\right)$
limits the possible effects in $\mathcal{B}\left(K_L\rightarrow\pi^0\nu \bar{\nu}\right)$, 
as expressed by the Grossman-Nir bound\cite{GN}, $\mathcal{B}%
\left(  K_{L}\rightarrow\pi^{0}\nu\bar{\nu}\right)  \leq4.4\times
\mathcal{B}\left(  K^{+}\rightarrow\pi^{+}\nu\bar{\nu}\right)  $,
corresponding to $\leq1.7\cdot10^{-9}$ (90\%), about 50 times the SM prediction.

Many models have been considered along the years, like for example 
the enhanced EW penguins\cite{BFRS04}, Little Higgs\cite{LH}, 
Extra dimensions\cite{ED},... which are encoded into $V\pm A$ FCNC operators, 
or leptoquark interactions\cite{LQ}, R-parity violating SUSY\cite{GN,Rparity},... 
which can give rise also to new scalar/tensor FCNC interactions. We will here 
concentrate on the MSSM.

\textbf{MSSM at large tan}$\beta$\textbf{:} When $\tan\beta=v_{u}/v_{d}\approx50$
say, the Higgs couplings to quarks get significant higher-order loop effects.
Of interest for $K\rightarrow\pi\nu\bar{\nu}$ is the charged Higgs contribution
to the $Z$ penguin\cite{isidoriP06}, which exhibits a $\tan^{4}\beta$ behavior,
is sensitive to $\delta_{RR}^{D}$ and is slowly decoupling when 
$m_{H^{\pm}}\rightarrow\infty$.

\begin{figure}[t]
\vspace{5pt} $
\begin{array}
[c]{c}%
\text{\includegraphics[width=65mm]{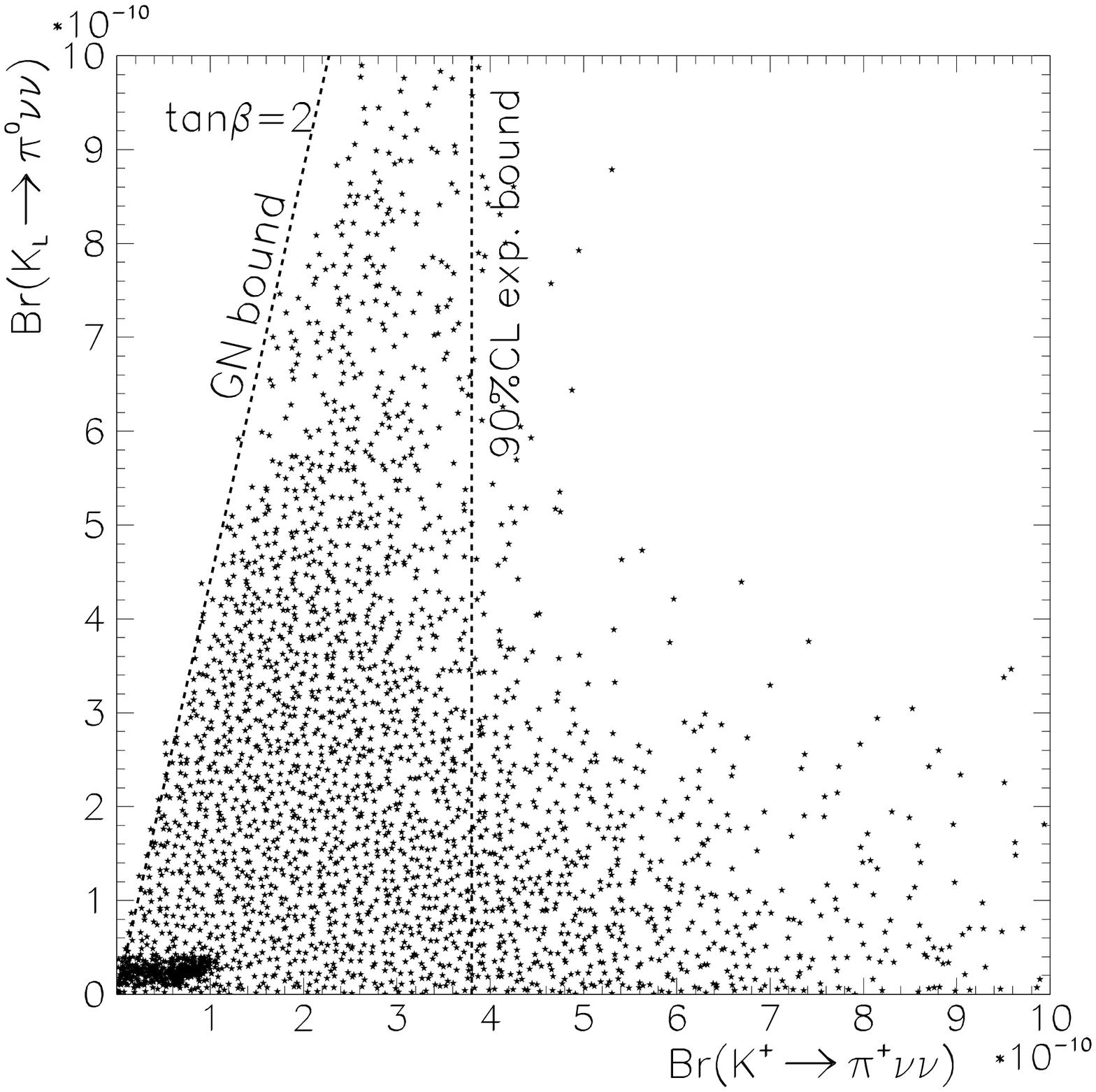}  }\\
\text{\includegraphics[width=72mm]{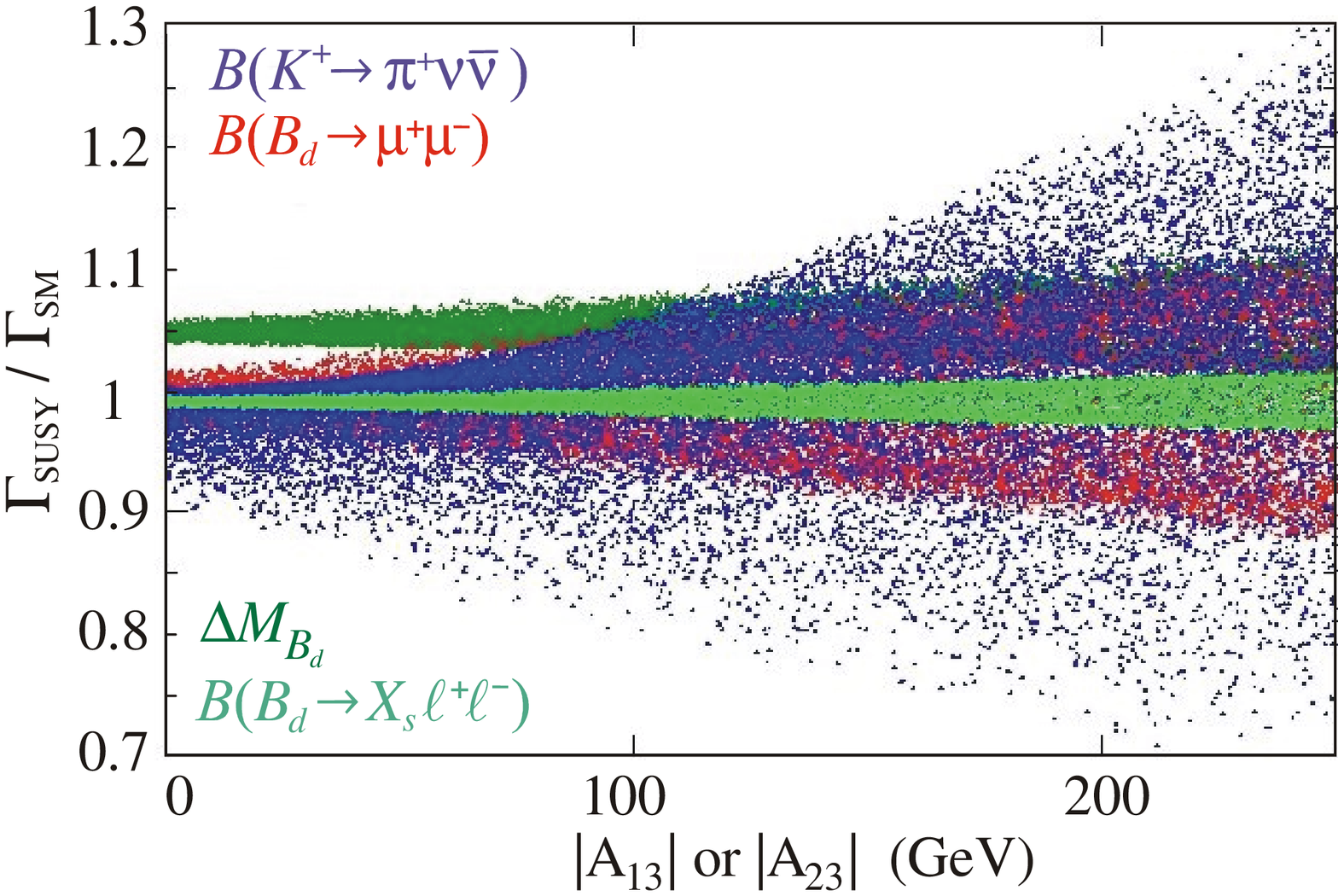}  }%
\end{array}
$\vspace{-0.6cm}\caption{Scan results of \cite{MSSM1}(up) and \cite{MSSM2}
(down).\vspace{-0.6cm}}%
\label{FigNP1}%
\end{figure}

\textbf{MSSM at moderate tan}$\beta$\textbf{:} In this case, the chargino
penguins are dominant\cite{Chargino}. Also, the single mass insertion approximation
is not sufficient, and these contributions probe the double $(\delta_{RL}^{U}%
)_{32}^{\ast}(\delta_{RL}^{U})_{31}$ insertion. Many works have analysed the
phenomenological consequences of these effects. Let us concentrate on two
questions of relevance for future experiments.

First, how large the effect on $\mathcal{B}\left(  K_{L}\rightarrow\pi^{0}%
\nu\bar{\nu}\right)  $ can be, given the current measurement of $\mathcal{B}%
\left(  K^{+}\rightarrow\pi^{+}\nu\bar{\nu}\right)  $. This has been answered
in \cite{MSSM1}, which showed that the GN bound can still be saturated in the
MSSM. A full scan over the parameters was performed, using adaptive numerical
algorithms (fig.\ref{FigNP1}, top).

A second question, especially relevant after a SUSY discovery at LHC, is 
how does the constraint from $K\rightarrow\pi\nu\bar{\nu}$ on the
trilinear terms $A^{U}$ compare with those from other K and B physics
observables. This has been answered in \cite{MSSM2}. Fig.\ref{FigNP1} (bottom)
shows that the $K\rightarrow\pi\nu\bar{\nu}$ decays are the most sensitive probe
of that sector.

\textbf{Minimal Flavor Violation: }If the SM Yukawas remain the only source of
flavor-symmetry breaking also beyond the SM, the FCNC remain tuned essentially
by the CKM matrix, hence are suppressed\cite{MFV}. This hypothesis can be
enforced model-independently, or, e.g., within the MSSM. In this latter case,
since the $t$-quark Yukawa is large, sizeable trilinear $A^{U}$ terms are still
allowed. As said previously, the $K\rightarrow\pi\nu\bar{\nu}$ modes are very
sensitive to that sector.

Still, MFV does its job perfectly in killing any large deviation with respect
to the SM. Though the MFV analyses in the literature differ in their
parametrization, statistical treatment of errors, extraction of CKM elements
and in the resulting correlations among observables, they all agree that the
enhancement of $K\rightarrow\pi\nu\bar{\nu}$ never exceeds
25\%\cite{MSSM2,MFVapp}.

\subsection{New Physics in $K_{L}\rightarrow\pi^{0}\ell^{+}\ell^{-}$ decays}

The $K_{L}\rightarrow\pi^{0}\ell^{+}\ell^{-}$ pair of decays is interesting
at least for three reasons. First, compared to $K\rightarrow\pi\nu\bar{\nu}$,
they can probe helicity-suppressed operators. Second, compared to $K_{L}\rightarrow
\mu^{+}\mu^{-}$, the theoretical control on the SM part is better and further,
$K_{L}\rightarrow\mu^{+}\mu^{-}$ is not sensitive to tensor operators.
Finally, $K_{L}\rightarrow\pi^{0}e^{+}e^{-}$ and $K_{L}\rightarrow\pi^{0}%
\mu^{+}\mu^{-}$ are two modes with very similar dynamics, but for the very
different lepton masses. This makes them ideal to probe NP effects through
their signatures in the pair\cite{MesciaST}.

\begin{figure}[t]
\vspace{5pt}\includegraphics[width=65mm]{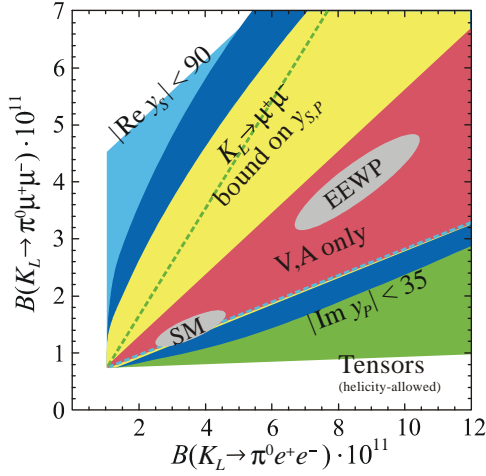}  \vspace{-0.6cm}%
\caption{$\mathcal{B}\left( K_{L}\rightarrow\pi^{0}\mu^{+}\mu^{-}\right)  $
against $\mathcal{B}\left( K_{L}\rightarrow\pi^{0}e^{+}e^{-}\right)  $ for
various NP scenarios\cite{MesciaST}.\vspace{-0.6cm}}%
\label{FigNP2}%
\end{figure}

\textbf{Vector and axial-vector operators: }The $(\bar{s}d)_{V}(\bar{\ell}%
\ell)_{V,A}$ operators, already present in the SM, arise for example from
EEWP\cite{BFRS04}, MSSM,... In general, these models also affect 
$K\rightarrow\pi\nu\bar{\nu}$, and the sensitivity is slightly
lower for $K_{L}\rightarrow\pi^{0}\ell^{+}\ell^{-}$ than for $K_{L}%
\rightarrow\pi^{0}\nu\bar{\nu}$. Anyway, they should not be disregarded
because, contrary to the neutrinos, they offer the possibility to disentangle
NP effects in $(\bar{s}d)_{V}(\bar{\ell}\ell)_{V}$ and $(\bar
{s}d)_{V}(\bar{\ell}\ell)_{A}$. Indeed, $(\bar{s}d)_{V}(\bar{\ell}\ell)_{A}$
produces the final lepton pair also in a helicity-suppressed $0^{-+}$ state,
hence contributes differently to $K_{L}\rightarrow\pi^{0}e^{+}e^{-}$ and
$K_{L}\rightarrow\pi^{0}\mu^{+}\mu^{-}$, while the $(\bar{s}d)_{V}(\bar{\ell
}\ell)_{V}$ contribution is identical (up to phase-space corrections).

This is depicted by the red region in fig.\ref{FigNP2}, which corresponds to
the region in the $\mathcal{B}\left( K_{L}\rightarrow\pi^{0}e^{+}%
e^{-}\right) -\mathcal{B}\left( K_{L}\rightarrow\pi^{0}\mu^{+}\mu
^{-}\right) $ plane spanned leaving $(\bar{s}d)_{V}(\bar{\ell}\ell)_{A}$ and
$(\bar{s}d)_{V}(\bar{\ell}\ell)_{V}$ operator coefficients arbitrary (but
keeping lepton universality). Taking all the errors into account, this
translates into the bounds $0.1+0.24\mathcal{B}^{ee}\leq\mathcal{B}^{\mu\mu
}\leq0.6+0.58\mathcal{B}^{ee}$ with $\mathcal{B}^{\ell\ell}=\mathcal{B}\left(
K_{L}\rightarrow\pi^{0}\ell^{+}\ell^{-}\right)  \cdot10^{11}$.

Finally, the contribution from EMO operator $(\bar{s}\sigma^{\mu\nu}%
d)F_{\mu\nu}$ can always be absorbed into a redefinition of $(\bar{s}%
d)_{V}(\bar{\ell}\ell)_{V}$\cite{BurasCIRS99}, and thus possible NP
contributions to it cannot be disentangled.

\textbf{Scalar and pseudoscalar operators}, $(\bar{s}d)_{S}(\bar{\ell}%
\ell)_{S\left(  P\right)  }$, induce a CP-conserving (CP-violating)
contribution, respectively. When these operators are helicity-suppressed, only
the muon mode is significantly affected. Such a situation corresponds for
example to the MSSM at large $\tan\beta$, where they arise from neutral Higgs
penguins and are sensitive to down-squark mass insertions\cite{IsidoriRetico}.
Combined with general $V,A$ operators, the blue regions in fig.\ref{FigNP2}
can be spanned.

Specific models like the MSSM can generate both $(\bar{s}d)_{S}(\bar{\ell}%
\ell)_{S,P}$ and $(\bar{s}d)_{P}(\bar{\ell}\ell)_{S,P}$ operators,
contributing to $K_{L}\rightarrow\ell^{+}\ell^{-}$. Working out their
relation, the current $\mathcal{B}\left( K_{L}\rightarrow\mu^+\mu
^- \right) ^{\exp}$ corresponds to the yellow region in fig.\ref{FigNP2}.

If the (pseudo-)scalar operators are helicity-allowed, the electron mode
becomes more sensitive, simply because of the phase-space suppression. Such
types of operators can arise from leptoquark tree-level exchanges\cite{LQ} or
sneutrino exchanges in SUSY without $R$-parity\cite{Rparity}. Still, operators
contributing to $K_{L}\rightarrow e^{+}e^{-}$ will also be generated. Such
contributions to an otherwise helicity-suppressed mode are very constrained by
$\mathcal{B}\left(  K_{L}\rightarrow e^{+}e^{-}\right) ^{\exp}=9_{-4}%
^{+6}\cdot10^{-12}$, and should not lead to a perceptible impact in fig.\ref{FigNP2}.

\textbf{Tensor and pseudotensor operators},\textbf{ }$(\bar{s}\sigma_{\mu\nu
}d)(\bar{\ell}\sigma^{\mu\nu}\ell)$ and $(\bar{s}\sigma_{\mu\nu}d)(\bar{\ell
}\sigma^{\mu\nu}\gamma_{5}\ell)$ induce a CP-violating (CP-conserving)
contribution, respectively.\textbf{ }In case these operators are
helicity-suppressed, being in addition phase-space suppressed, their impact on
$K_{L}\rightarrow\pi^{0}\mu^{+}\mu^{-}$ is smaller than for scalar and
pseudoscalar operators. In addition, in models like the MSSM, they are further
suppressed by loop factors\cite{BBKU02} and their impact can be expected to be small.

On the other hand, if helicity-allowed, there are at present no constraint on
them, since they do not contribute to $K_{L}\rightarrow\ell^{+}\ell^{-}$. This
is depicted by the green region in fig.\ref{FigNP2}.

\section{Conclusion}

Rare K decays are very clean and sensitive probes of New Physics. They are promising
not only to eventually get clear signals, but also to constrain the nature of
the New Physics at play through the pattern of deviations they could exhibit
with respect to the SM predictions.

\end{document}